\documentclass{webofc}
\usepackage{eurosym}
\begin{document}
\title{The case for an all-sky millimetre survey at sub-arcminute resolution}
%
% subtitle is optionnal
%
%%%\subtitle{Do you have a subtitle?\\ If so, write it here}

\author{\lastname{François-Xavier Désert}\inst{1}\fnsep\thanks{\email{francois-xavier.desert@univ-grenoble-alpes.fr}} \and
        \lastname{Martino Calvo}\inst{2} \and
        \lastname{Andrea Catalano}\inst{3} \and
        \lastname{Samuel Leclercq}\inst{4} \and
        \lastname{Juan Macias-Perez}\inst{3} \and
        \lastname{Frédéric Mayet}\inst{3} \and
        \lastname{Alessandro Monfardini}\inst{2} \and
        \lastname{Laurence Perotto}\inst{3} \and 
        \lastname{Nicolas Ponthieu}\inst{1}
}

\institute{Univ. Grenoble Alpes, CNRS, IPAG, 38000 Grenoble, France
\and
           Univ. Grenoble Alpes, CNRS, Grenoble INP, Institut Néel, 38000 Grenoble, France
\and
           Univ. Grenoble Alpes, CNRS, LPSC/IN2P3, 38000 Grenoble, France
           \and
           Institut de RadioAstronomie Millimétrique (IRAM), 38000 Grenoble, France
          }

\abstract{%
   There are several new projects to survey the sky with millimetre eyes, the biggest being Simons Observatory and CMB-S4, in the Southern Hemisphere. The NIKA2 collaboration has acquired sufficient knowledge to build a large focal plane KID camera for a 15~m antenna. This would allow covering the whole Northern Hemisphere in five years at subarcminute resolution and with milliJansky point-source sensitivity. We describe the main scientific drivers for such a project: the SZ sky, the high-redshift millimetre Universe and the interstellar medium in our Galaxy and the nearby galaxies. We also show briefly the main difficulties (scientific, organisational, technical and financial).
}
\maketitle
\section{Introduction}
\label{intro}
I would like to present you with the Grenoble project that we are designing in order to make a millimetre survey of the whole sky  below the arcminute resolution. This project is not yet funded. At this stage, we are seeking feedback from the community to see if there is a general interest. Clearly, we come from the NIKA2 legacy. We have learned many lessons from NIKA2 at the IRAM 30~meter telescope both in terms of scientific outputs and in astronomical instrumentation  \cite{Perotto2020} , and we can put this knowledge to good use for this new project, called HR15m. HR15m is a millimetre telescope of 15\,m diameter with a single photometric mapping instrument covering the atmospheric windows from 1 to 3.3\,mm at 20 to 60\,arcsecond diffraction-limited resolution. Equipped with about 20,000\,KIDs at 150\,mK over a Field-Of-View diameter of 48\,arcminutes, the instrument would have a ten-fold increase in mapping speed with respect to NIKA2, and it would survey the whole Northern Hemisphere in five years.

\section{Scientific goals}
\label{Sci}
\subsection{All the milliJansky point-sources of the Northern Sky}
\begin{figure}[ht]

\centering

\includegraphics[scale=0.3]{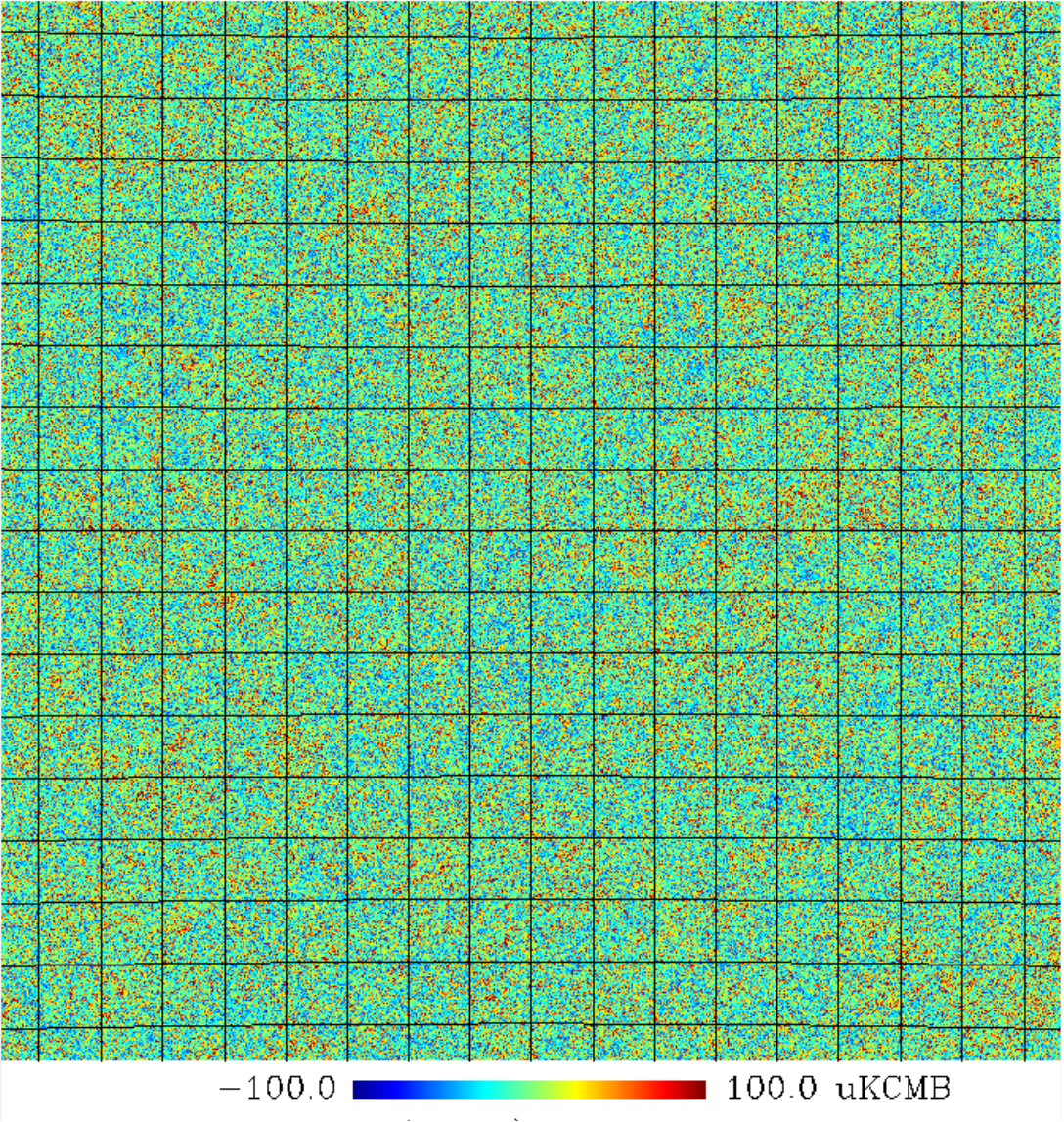}
\caption{CIB simulation of a patch of sky the size of HR15m deep survey. Each square has a 1~degree side.}
\label{fig-1}       % Give a unique label
\end{figure}

What would we observe with such a telescope? Clearly a deep survey is needed. We show in Figure~\ref{fig-1} the simulation of a 17~deg. wide square map, which would be a typical deep survey of 300 square degrees. The simulation is performed by using WebSky tools \cite{Stein2020} at one millimetre wavelength, and we are dominated by the Cosmic Infrared Background (CIB). So we will get a survey of that zone at the 1~mJy $5\,\sigma$ level. 

Using models of the number counts that are matching the different current observations (\cite{Bethermin2017}, \cite{Vieira2010}, \cite{Tucci2011}), we can evaluate the number of sources that could be detected in such a deep survey: 120,000 sources above $5\,\sigma$~(1.8 mJy at 260~GHz) which has to be compared with the few hundreds of sources that we get with NIKA2 deep surveys \cite{Bing2023} . The confusion would be reached at 300,000 sources. So we are near the confusion limit of a 15~m telescope, and we can also perform wide surveys, say six thousand square degrees and in that case we are aiming at finding rare objects which are probably lensed images of very distant galaxies. An expected number of lensed objects above 10~mJy is around 1300 at 220~GHz. This can be compared with Herschel successful surveys of ten to one hundred lensed objects. Many radio sources would also be found: 15,000 at 150~GHz and 27,000 at 90~GHz.
One of the main goals of HR15m is therefore to be a CIB millimetre mapper. The use of these sources covers correlation functions, but also CMB delensing and masking in CMB-SO/S4 projects~\cite{SO_2019},~\cite{S4_2019} (in the common area with the coverage of these Southern telescopes).

\subsection{All the massive clusters in the Northern Hemisphere}

\begin{figure}[ht]
\centering
\includegraphics[scale=0.3]{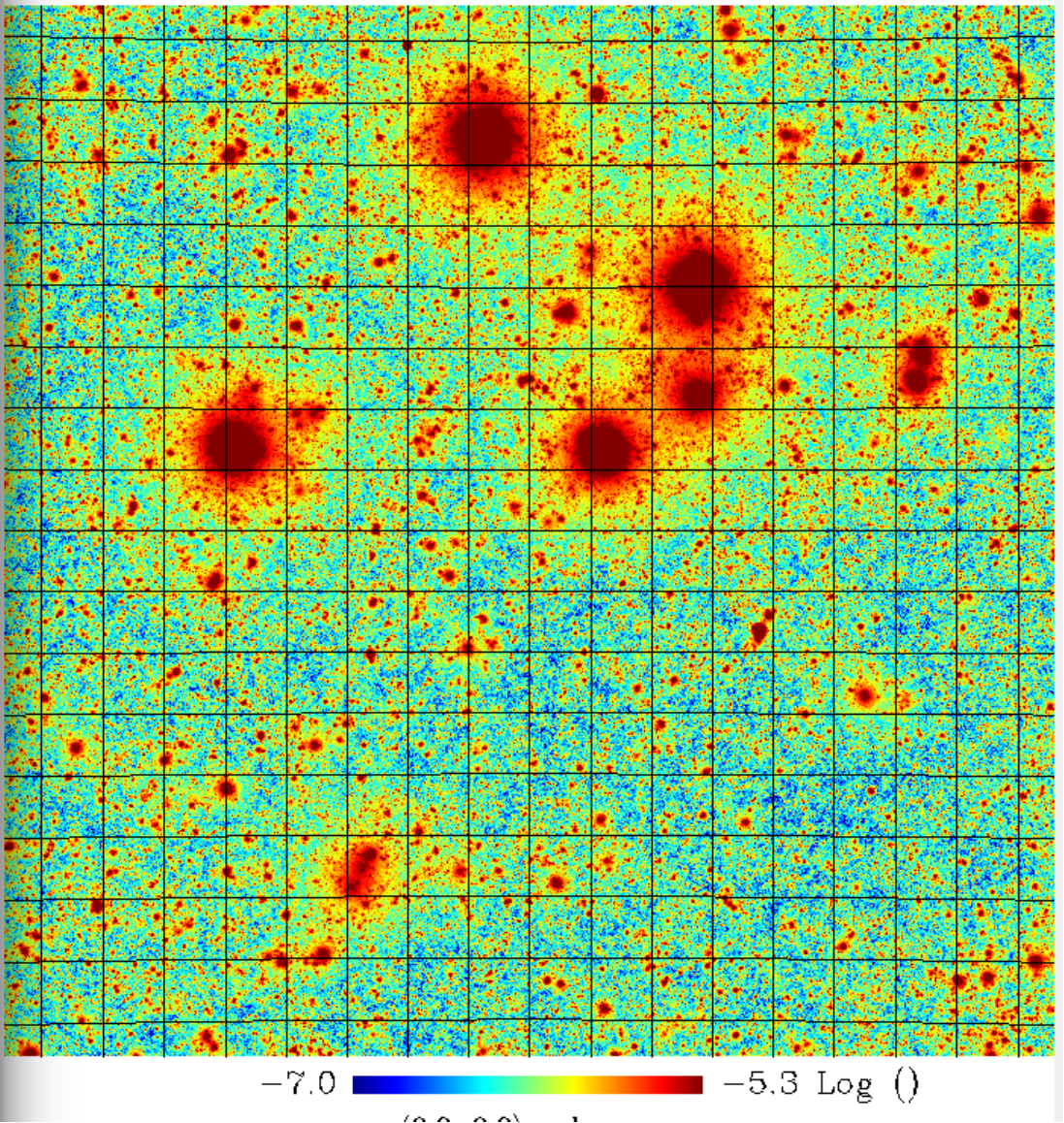}
\includegraphics[scale=0.3]{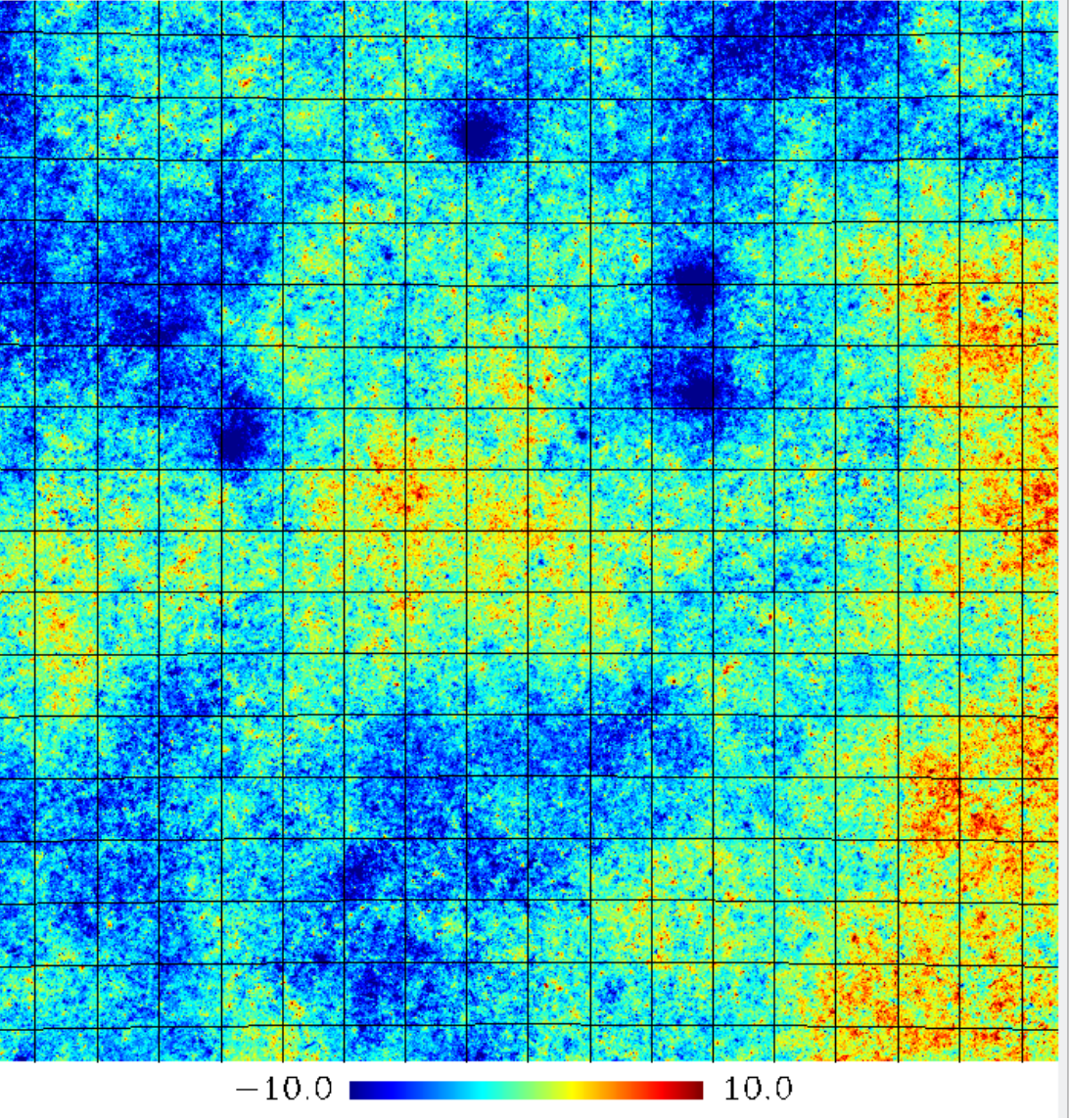}
\caption{tSZ and kSZ maps of the same patch of sky, corresponding to a deep survey. These are WebSky simulations.}
\label{fig-2}       % Give a unique label
\end{figure}

Starting from the Planck tSZ (thermal Sunyaev-Zel'dovich effect) source counts~\cite{Planck2014} and assuming that clusters have a diameter of at least 2~arcmin, the wide survey would deliver 26,000 clusters at $5\,\sigma$ with an integrated Compton parameter above $8\times 10^{-5}\,\mathrm{arcmin^2}$ at 150~GHz. The deep survey would show (coincidentally) the same number of clusters but with $Y_{SZ} \ge 2\times 10^{-5}\,\mathrm{arcmin^2}$ (see Fig.~\ref{fig-2}).

\begin{figure}
% Use the relevant command for your figure-insertion program
% to insert the figure file.
\centering
%\sidecaption
\includegraphics[scale=0.2]{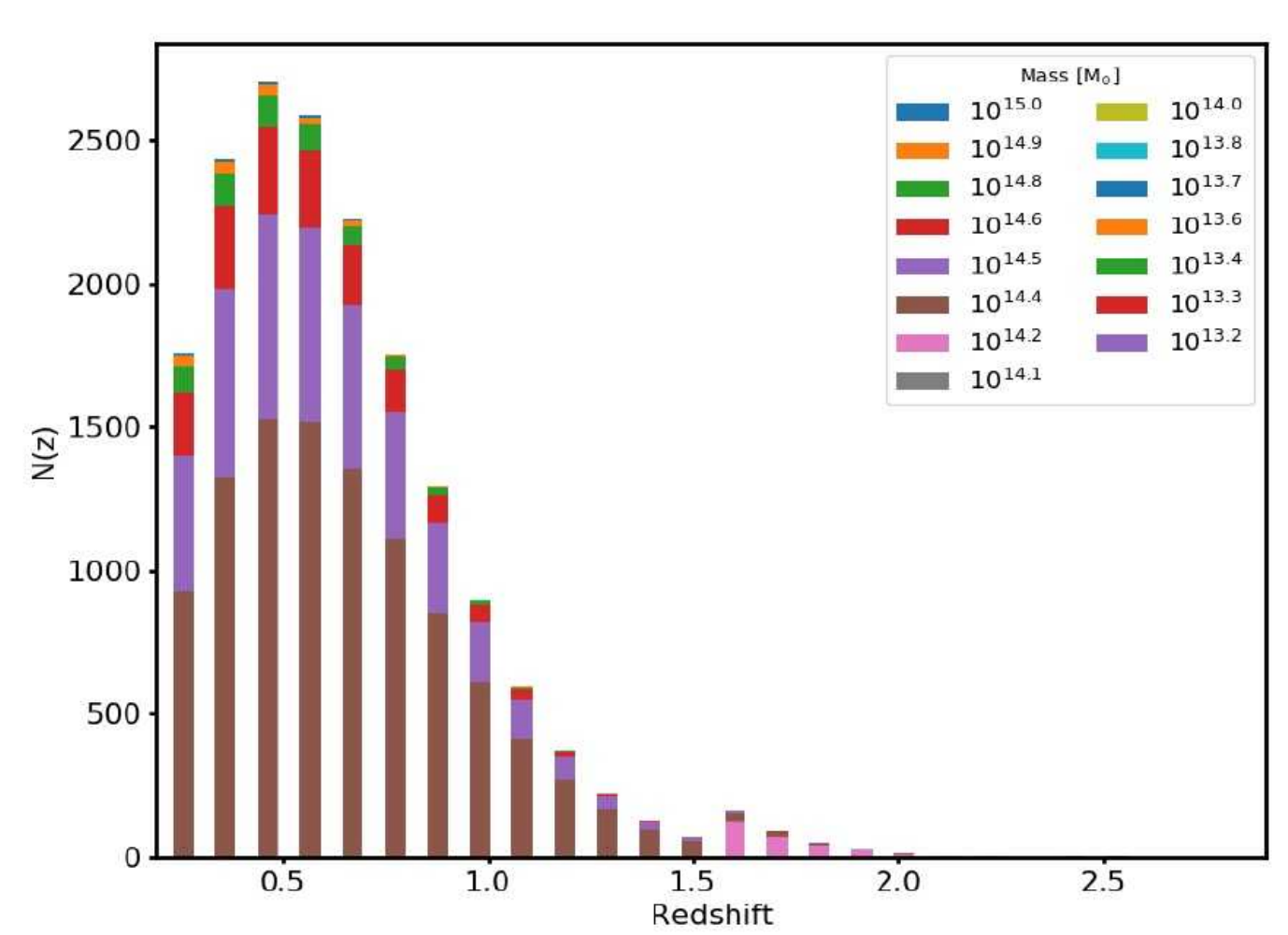}
\includegraphics[scale=0.2]{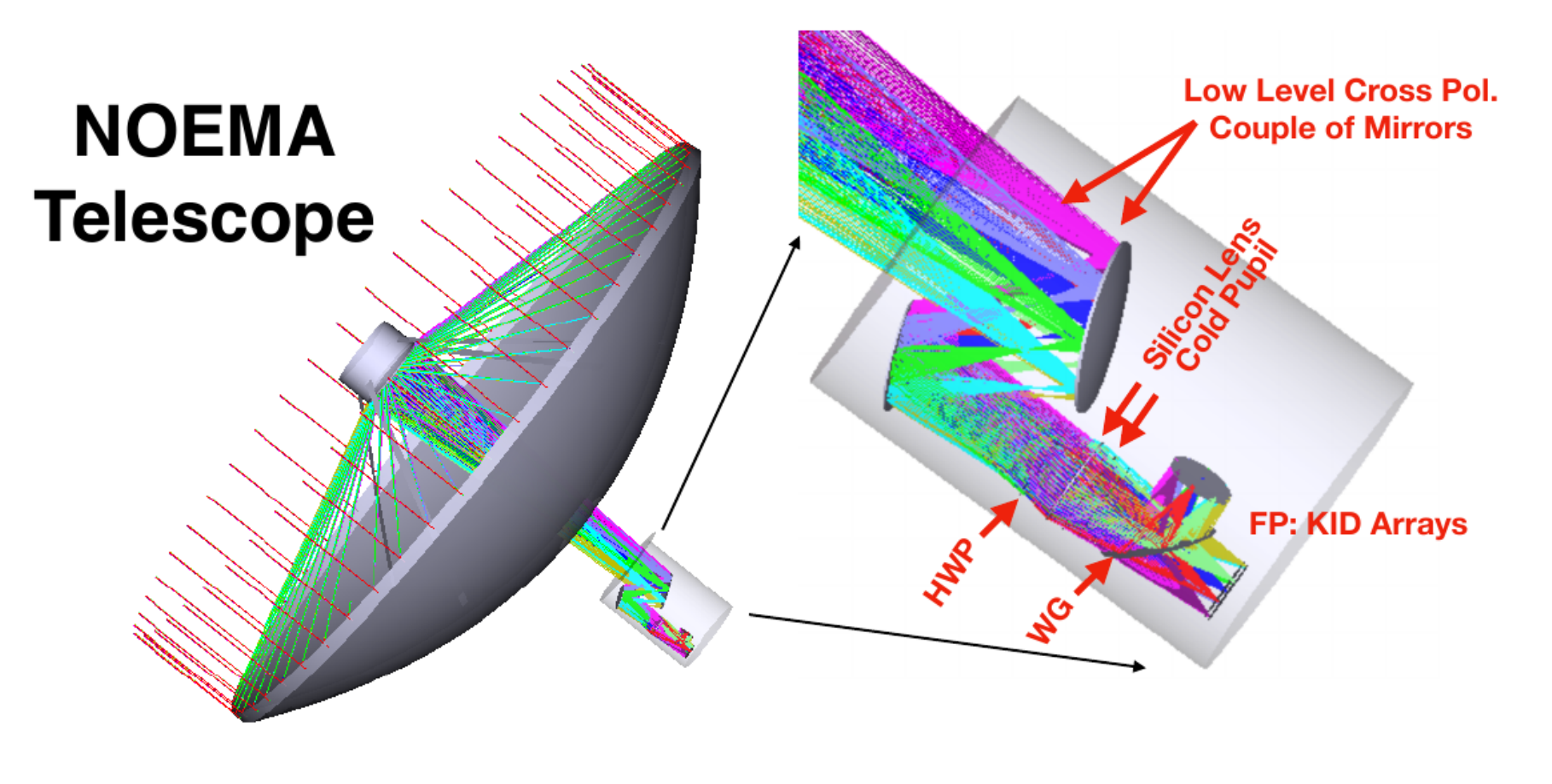}
\caption{Left: redshift distribution of detected clusters of galaxies according to their mass, in the wide survey case. Right: a preliminary optical implementation of HR15m.}
\label{fig-3}       % Give a unique label
\end{figure}

The histogram of the redshift of detected clusters, shown in the left panel of Fig.~\ref{fig-3}, reveals that we can observe many high-redshift clusters. This would provide direct access to cluster formation and evolution. The surveys would help to map filaments in the Cosmic Web, and in particular the missing baryons~\cite{Kaastra2013}. It would show the cosmic matter out of which clusters grow. In other words, we can observe how kinetic energy is transformed into thermal energy by using the kSZ (see below) and  the tSZ effect.

\subsection{A window to the reionization epoch}

We can even observe routinely the kinetic SZ (kSZ) effect, which has the same colour as CMB anisotropies, but which appears at much smaller angular scales. In order to null the main thermal effect (tSZ), we combine the different photometric bands of HR15m in order to get a kSZ sensitivity (after component separation) of $9\,\mu\mathrm{ K_{CMB}}$, $1\,\sigma$, in one arcminute pixel. This sensitivity is rather limited for an individual cluster, but statistics of clusters will help to detect motions in the large-scale structure of the Universe. Once the CIB can be disentangled from the kSZ \cite{Addison2012}, one can have access to kSZ traces of the reionization period~\cite{Douspis2022}, which can be cross-correlated with SKA maps at small angular scales.

More subtle effects can even be looked for, owing to the large number of detected clusters. CMB lensing by clusters (\cite{Hu2007}, \cite{Lewis2006}) could be detected using HR15m, if many clusters are stacked along the gradient of the CMB. The gravitational effect of clusters is to invert locally the temperature gradient and produce a typical local dipole signal. It should prove an interesting  statistical measure of the lensing by clusters on the CMB. We estimate the significance at $10\,\sigma$ level when stacking 1000 clusters. This could be another mass-proxy for clusters of galaxies.

\section{The project characteristics}

\subsection{Sky coverage}
\label{sub-sky}

\begin{figure}[ht]
% Use the relevant command for your figure-insertion program
% to insert the figure file.
\centering
%\sidecaption
\includegraphics[width=11cm,clip]{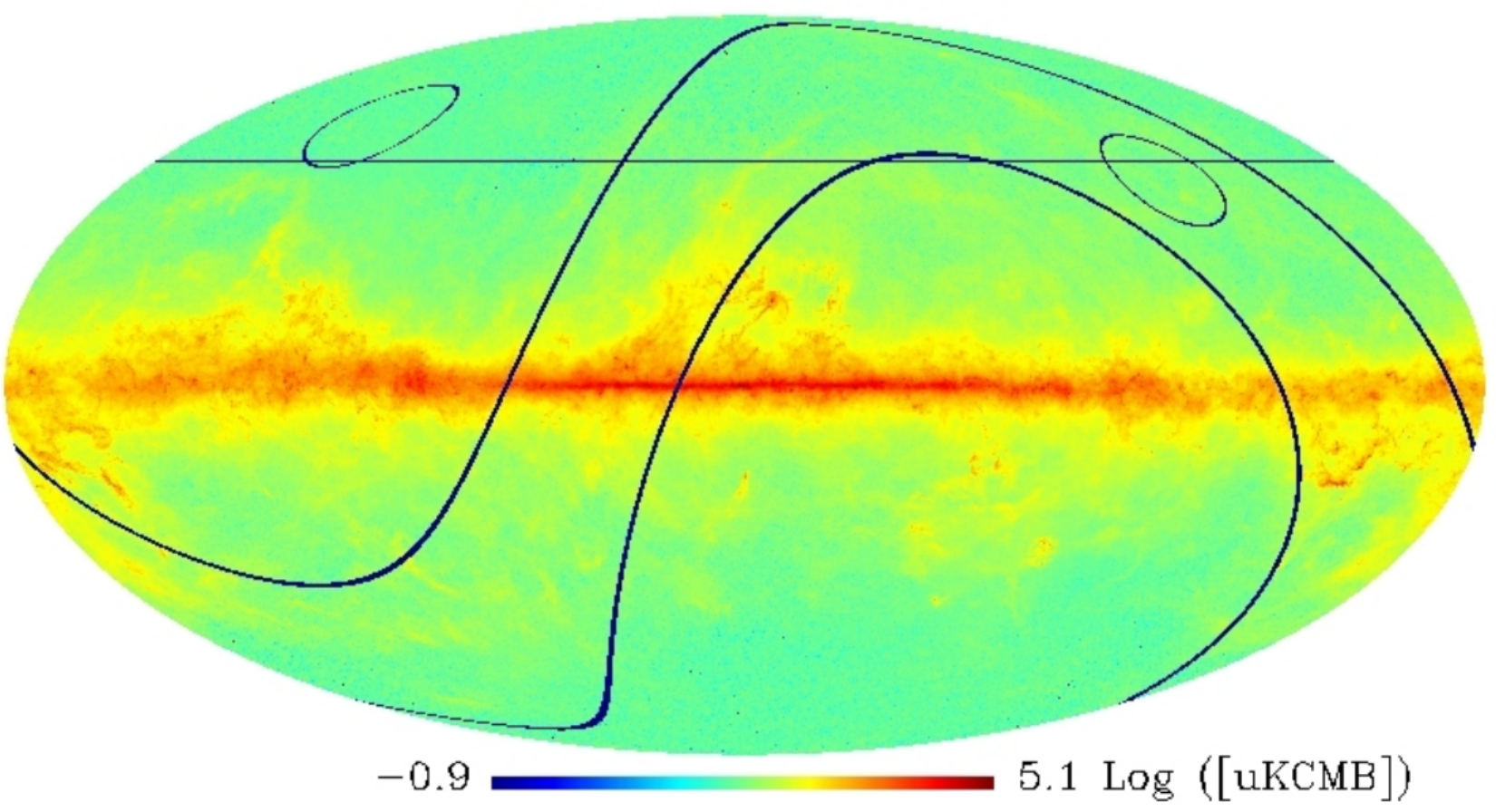}
\caption{HR15m sky coverage. Mollweide all-sky map showing the Planck NILC 217~GHz emission, without the CMB anisotropies, in log scales and galactic coordinates. The possible deep survey areas are shown as the two ellipses (centred on GoodsNorth - left ellipse- and Cosmos - right ellipse - fields). One possible wide survey is the Northern Galactic Cap as limited by the horizontal line. The Northern Sky (with a declination lower limit of -15\,deg.) is above the nearly circular thick black line on the right, whereas the Southern Hemisphere telescopes have access to the sky below the left S-shaped line. The overlap zone can be used for cross-correlating results in both hemispheres}
\label{fig-4}       % Give a unique label
\end{figure}

Fig.~\ref{fig-4} shows the sky coverage of HR15m. This project will map the whole Northern sky. ACT has partially mapped the Southern sky \cite{Naess2020} albeit with a lower angular resolution (2') and the CMB-SO/S4 projects will complete that task in the Southern Hemisphere, but there is no project to do that in the Northern Hemisphere. 

In parallel with the two deep surveys, of $300\,\mathrm{deg.^2}$ size each, in 2 years, the telescope can cover the whole Northern sky ($\delta \ge -15\,\mathrm{deg.}$) in 4 years ($6000\,\mathrm{deg.^2}$ per year). Realistic sensitivities, used in Sect.~\ref{Sci}, were computed by assuming an average 4\,mm of precipitable water vapour, an average 45\,degree elevation, an instrument efficiency of 80\,\%, a 50\,\% weather efficiency, and by enabling calibration during a third of the time on sky. 

\subsection{Telescope}

We propose here the implementation of a NOEMA-like 15~m antenna as shown in Fig.~\ref{fig-3} (right panel). You can see the cryostat at the Cassegrain focus with an early implementation, where we considered the polarization option (this option could be considered as an upgrade of the basic intensity-only configuration envisaged at the start). The optics, in the case of pure photometry, would be a bit simpler, and we can accommodate a field-of-view diameter of typically 48~arcminutes, where the instrument would split the different wavelengths with dichroics. 

This design is very preliminary, but it shows that covering a large field of view with an off-the-shelf 15-metre telescope is feasible.

One of the constraints that we have identified comes from the pointing requirements to beat the sky noise and the electronic noise. We need to be able to scan the sky as fast as possible and, because Kids have negligible time constant, we can scan the sky typically at 10~arcminutes (about two Planck 1~mm beams) per second which is one field of view every four seconds. The data rate of 1~TB per day would not be a major problem because the Concerto team has shown how to handle that \cite{Concerto2020}  . Large angular scale HR15m data will be smoothly joined with the Planck survey (see {\it e.g.}~\cite{Naess2020}).

\subsection{Instrumental requirements}

The successful legacy of NIKA2~\cite{Adam2018} and Concerto~\cite{Concerto2020} allows us to be confident in the swift production of new KID arrays. The major hurdle here is the scaling of the number of KIDs by a factor of 10 with respect to the existing instruments. HR15m requires in the order of 20,000 KIDs. The authors belong to a group of Grenoble laboratories (GIS KID) that has a long history of taking innovative detectors to the telescopes. The basic LEKID design is with Aluminium or TiAl. We need to improve on uniformity of detectors for 10-kilo pixel arrays. We plan on using 4-inch wafers of monochromatic detectors: a new machine is ordered in order to fabricate a whole array on a single wafer. The readout units are designed with a 1~GHz bandwidth in order to accommodate a frequency multiplexing factor of at least 500.

The HR15m bandpass requirements come from the main scientific goals enumerated in Sect.~\ref{Sci} which demand spectral diversity. We need five bandpasses covering up to a 50~GHz width, avoiding the atmospheric oxygen and water lines and centred at those frequencies: 90, 150, 220, 250 and 280~GHz, in a very similar pattern to the one of CCAT-Prime~\cite{Choi2020}.

\subsection{Budget}

A rough-order-of-magnitude costing leads to 5-10~M\euro{} for the telescope (a modified copy of a NOEMA dish), 5-10~M\euro{} for the instrument and 1-2~M\euro{} for the annual running cost (times six years of commissioning and surveys). The size of this project clearly requires a European collaboration based on an existing Observatory: IRAM Plateau de Bure, IRAM Pico Veleta, Teide Observatory, or at least a millimetre-quality site~\cite{Raymond2021}.
The medium cost of HR15m has to be compared with the financing of the millimetre giants (SPT, ACT, ALMA, CCAT-Prime~\cite{Choi2020}, CMB-HD~\cite{Sehgal2019}, Planck, Litebird) which are all in the several hundred M\euro{} range or more. The human organisation would be of the size of the NIKA2 collaboration (50–100 scientists) but a survey mode will be simpler to manage than open-time observations. A possible optimistic timing would consist in the design and construction between 2024-2028, commissioning in 2029 and survey in 2030-2034.

\section{Conclusion}

HR15m is a medium cost, fast-track, all sky millimetre survey telescope at high angular resolution: where ISM meets CMB! It is the logical sequel to Planck millimetre survey legacy.
The main drivers are the CIB anisotropies, the radio sky, the tSZ and kSZ sky. Nearby galaxies and the ISM will be automatically observed. Planet 9 ({\it e.g.}~\cite{Naess2021}) and transient objects are in the big realm of serendipity offered by this project.
HR15m will be an important complement to the CMB-S4 efforts, with its higher angular resolution providing added value to the SZ science.
The most interesting sources will have to be investigated further with other facilities like NOEMA and ALMA. The best use of the produced homogeneous dataset will be through cross-correlation studies with the wealthy datasets from modern ground-based and space experiments. The question of the Southern Hemisphere will be open once we see the outcome of the Large Angular Telescopes which are part of the SO and S4 projects.

\section{Acknowledgements}
We acknowledge the use of the Legacy Archive for Microwave Background Data Analysis (LAMBDA), part of the High Energy Astrophysics Science Archive Center (HEASARC). HEASARC/LAMBDA is a service of the Astrophysics Science Division at the NASA Goddard Space Flight Center.
This work is based on observations obtained with Planck (http://www.esa.int/Planck), an ESA science mission with instruments and contributions directly funded by ESA Member States, NASA, and Canada.
We thank many discussions with members of the NIKA2 and CONCERTO collaborations.
%For bibliography use \cite{Keruzore:2022tpj}
%paragraph a unique label (see Sect.~\ref{sec-1}).

%For two figures side-by-side use syntax of figure~\ref{fig-2} and adjust the scale factor.

%For figure with sidecaption legend use syntax of figure~\ref{fig-3}.


\begin{thebibliography}{}
\bibitem[]{Adam2018}
R.~Adam et al., Astron. Astrophys. \textbf{609}, A115 (2018)
\bibitem[]{Addison2012}
G. E. Addison et al., MNRAS \textbf{427}, 1741 (2012)
\bibitem[]{Bethermin2017}
M.~Béthermin et al., Astron. Astrophys. \textbf{607}, A89 (2017)
\bibitem[]{Bing2023}
L. Bing et al., Astron. Astrophys., \textbf{677}, A66 (2023)
\bibitem[]{Choi2020}
S.~K.~Choi et al., Journal of Low Temp. Phys. \textbf{199}, 1089 (2020)
\bibitem[]{Concerto2020}
Concerto Collab., P. Ade et al., Astron. Astrophys. \textbf{642}, A60 (2020)
\bibitem[]{Douspis2022}
M. Douspis et al., Astron. Astrophys. \textbf{659}, A99 (2022)
\bibitem[]{Hu2007}
W.~Hu et al., New J. of Phys. \textbf{9}, 441 (2007)
\bibitem[]{Kaastra2013}
J.~Kaastra et al., An Athena+ supporting paper, (2013), arXiv:1306.2324
\bibitem[]{Lewis2006}
A. Lewis \& A. Challinor, Phys. Rep. \textbf{429}, 1 (2006)
\bibitem[]{Naess2020}
S.~Naess  et al., Journal of Cosmology and Astroparticle Physics \textbf{12}, 04 (2020)
\bibitem[]{Naess2021}
S.~Naess  et al., Astrophys. J. \textbf{923}, 224 (2021)
\bibitem[]{Perotto2020}
L. Perotto et al., Astron. Astrophys. \textbf{637}, A71 (2020)
\bibitem[]{Planck2014}
Planck Collab., P.~A.~R.~Ade et al., Astron. Astrophys. \textbf{571}, A20 (2014)
\bibitem[]{Raymond2021}
A.~W.~Raymond et al., Astrophys. J. Supp. Ser. \textbf{253}, 5 (2021)
\bibitem[]{S4_2019}
J.~Carlstrom et al., Bull. of the Am. Astron. Society \textbf{51}, 7 (2019)
%save space, arxiv.1908.01062  
\bibitem[]{SO_2019}
P.~Ade et al., Journal of Cosmology and Astroparticle Physics \textbf{02}, 56 (2019)
\bibitem[]{Stein2020}
G.~Stein et al., Journal of Cosmology and Astroparticle Physics \textbf{10}, 12 (2020)
\bibitem[]{Sehgal2019}
N.~Sehgal et al., Bull. of the Am. Astron. Society \textbf{51} (2019), arXiv.1906.10134
\bibitem[]{Tucci2011}
M.~Tucci et al., Astron. Astrophys. \textbf{533}, A57 (2011)
\bibitem[]{Vieira2010}
J. D. Vieira et al., Astrophys. J. \textbf{719}, 763 (2010)

\end{thebibliography}
\end{document}